# Vehicle Intrusion And Theft Control System Using GSM and GPS;

An advance and viable approach

Ashad Mustafa, Hassan Jameel, Mohtashim Baqar, Rameez Ahmed Khan, Zeeshan M Yaqoob, Zeeshan Rahim & Safdar Khan

*Abstract*—This paper presents a novel approach towards the designing and development of a feasible and an embedded vehicle intrusion and theft control system using GSM (Global System for Mobile Communication) and GPS (Global Positioning System). The proposed system uses GSM technology as one of the distinguishing building blocks of the system. A GPS module Holux GR89 is used to trace the position of the vehicle and Mercury Switches are used to collect analog data continuously, in case of an intrusion, variations will be observed in sensors reading. Continuous readings from sensors are collected on to the microcontroller constantly and on the basis of those readings decision is taken whether an intrusion is made or not and in case of an intrusion a message from a predefined set of messages is generated to the owner of the vehicle and on reception of the message, the owner will have the luxury to take an action via an SMS either to lock the gears of the vehicle or seize the engine of the vehicle from a far-off place. A relay is used to control gears and engine of the vehicle while working with the microcontroller. A prototype system was built and tested. The results were very positive and encouraging.

*Keywords-component; gsm, microcontoller, gps, car intrusion, car theft, sensor.*

## I. INTRODUCTION

Over the past number of years real-time tracking and monitoring is one of the areas of escalating interest where a lot of work has been done and a lot of developments have been made to use it for the benefit of mankind. Now-a-days security is one of the key issue and basic need of our society. With fleeting time crime rates are also on the rise and will continue to increase as time passes. So, for present and future, security and authentication is the need of the hour [1]. A lot of progress has been made in this regard and most of the implementation can be seen in our daily lives in face of vehicle monitoring and tracking systems, weather forecasting, weapon detectors, heat sensing devices etc. This paper explains an effective and advanced system implementation of vehicle tracking and monitoring using GSM [2][3]. Due to a continuous rise in crime rates for the past no. of years, the conventional (alarm) vehicle monitoring and tracking systems used have fallen short to the mark. Some the reasons of lacking are listed below:

- Distance-the siren cannot be heard over long range of distance.
- Alarms can be made to disabled.
- Most of the cars have alarms with similar sound.
- Alarm sound is mitigated in crowded areas

The proposed system has been designed to work with GSM technology, which will generate a message, every time an intruder tries to get unauthorized access of the vehicle. GSM being one of the most popular and used mean of mobile communication makes it viable and unique in a way that many of the systems/applications designed can be made to work with GSM because it is a worldwide used, implemented and followed standards [4][5]. The proposed system, on intrusion, triggers the location of the vehicle to the owner via SMS and continues to do so with a constant interval of time. One of the distinguished features the system provides is that it allows the owner to prompt an action via SMS from a far off place like gear locking or cause an engine to seize etc.

The tracking feature has been incorporated using a GPS Tx/Rx module which serves the purpose of information gathering about the position of the vehicle. Mercury sensors are used for the purpose of sensing if anyone disturbs the stationary state of the vehicle it prompts microcontroller to cause an action. It gives continuous readings to the microcontroller on the basis of which decisions are taken. A set of predefined messages are stored each for a particular reading and to prompt a particular action.

## II. METHODOLOGY

The methodology adapted in the proposed model is that an interface of mercury switches, GPS module, and GSM module is being developed with a microcontroller. The communication between the microcontroller and other components of the system takes place serially. The microcontroller continuously receives data from the mercury switches and GPS module, takes decision on the basis of the readings collected onto the microcontroller. On intrusion the

Ashad Mustafa, is with the FEST, Iqra University Defence view campus Karachi, Pakistan; (e-mail: sashad@iqra.edu.pk).

Hassan Jameel, is with the FEST, Iqra University Defence view campus Karachi, Pakistan.

Mohtashim Baqar, is with the FEST, Iqra University Defence view campus Karachi, Pakistan.

microcontroller will generate a message to the vehicle owner or to a set of predefined numbers stating the current location of the vehicle and what sort of an intrusion has been made, has it been made with doors, bonnet or trunk. On reception of the message the user will have the option to prompt an action from a far-off place like locking the doors, seize the engine, or cut the supply line. A message from the owner will be collected on to the microcontroller via GSM module, Based on the message collected a relay will be switched to take an action like locking the doors, seize the engine or cutting the supply line.

### III. SYSTEM ARCHITECTURE

Following are the building blocks of the proposed model integrated together to make a complete system, a general block diagram of the project can be seen in the figure 1.

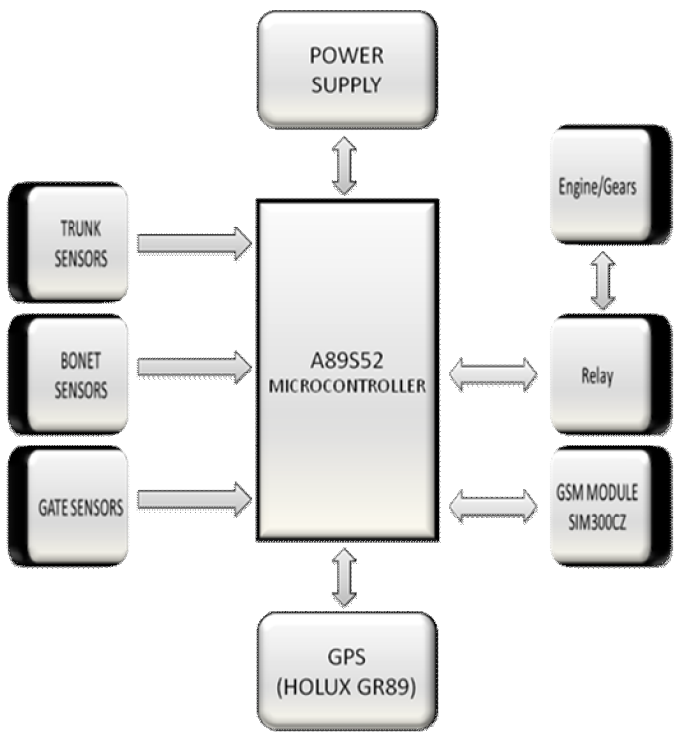

Fig 1: Block Diagram of the proposed model

#### A. GSM Module SIM300

SIM300CZ is a powerful tool which can work on frequencies EGSM 900 MHz, DCS 1800 and PCS 1900 MHz's. It is a tri-band GMS/GPRS device. It can also support GPRS coding schemes like CS-1, CS-2, CS-3 and CS-4. With small size it is suitable for industry requirement. For example in Machine-to-Machine, and mobile data communication systems SIM300CZ is best option. As it has an integrated charge circuit built-in; best option for circuits use batteries.

In this project SIM300CZ sends SMS to owner of car when microcontroller commands it to send SMS.

#### B. Microcontroller 8052

Microcontroller 8052 is a very significant part of this system. It is like PC on chip. Its basic features are;

It has 8KB ROM, 256 Bytes RAM, and 3 timers built-in. It is a cost effective and low power consumption device. With serial link connected to PC, it can be programmed directly or program can be uploaded after writing offline program on PC [6][7].

It receives alarm signal from sensors and after execution it sends command to SIM300 to send SMS.

#### C. GPS Holux GR89

To determine the exact location of car GPS module Holux GR89 has been used. It is a small size low weight and low power consumption device. Some technical details are as under;

It has 49 MHz processor, 20 Channel GPS receiver, and 200,000 effective correlators for fast Time to First Fix (TTFF), even at poor satellite signals. It Supports NMEA-0183 v2.2 data protocol: SiRF binary code. It has 4Mbit integrated program Flash and ARM7TDMI processor.

It sends location to GSM module to include in SMS via Microcontroller.

#### D. Sensors / Mercury Switches

There are so many sensors available in market. Mercury switches are being utilized for this project. Reason for using mercury switch was its working. Mercury switch is also called Mercury tilt switch. It works as; mercury is sealed in glass bulb or tube along with two or more electrical contacts. Mercury is constantly forced to go to the lowest point in bulb by gravity. When it flows towards electrical contacts and touches them and causes close switch. And when it is on other side of electrical contacts; causes open circuit. In vehicle this switch is deployed as a sensor in bonnet, doors and trunk. Whenever someone tries to open the vehicle door, bonnet or trunk, switch is placed in such a manner that it will close and send the alarm signal to Microprocessor.

#### E. Relays and Additional Circuits

Relays are being used to trigger an action; if an intrusion is being made the owner of the car can prompt an action from a far-off place through an SMS. Relays used works as a switching unit. One end of the relay is connected to the microcontroller while the other end is being connected to the battery, gear unit or the supply line to the engine. An action can be taken depending upon the user. Microcontroller will generate a signal depending upon the message signal send by the owner of the car to the microcontroller via GSM module.

Additional or supporting circuitry like power supply and snubber circuit is being used. Snubber circuit is used to control the surge of current that passes through the switch when it is closed and also to avoid the burning out of the switch when the circuit is open and all the voltages are acting on the terminal of the switches. The power required to the system is being drawn from the battery of the vehicle; a rheostat is being deployed in between the battery and the system to limit the current to the system to its requirement.

location update. For the communication purpose AT commands are used. Table 1 shows some of the commands implemented using this system.

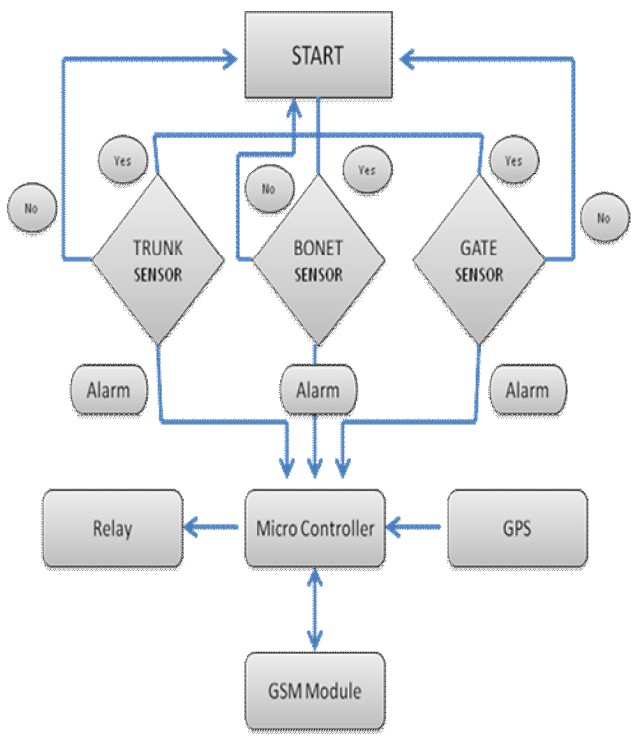

Fig 2: Flow chart of the proposed system

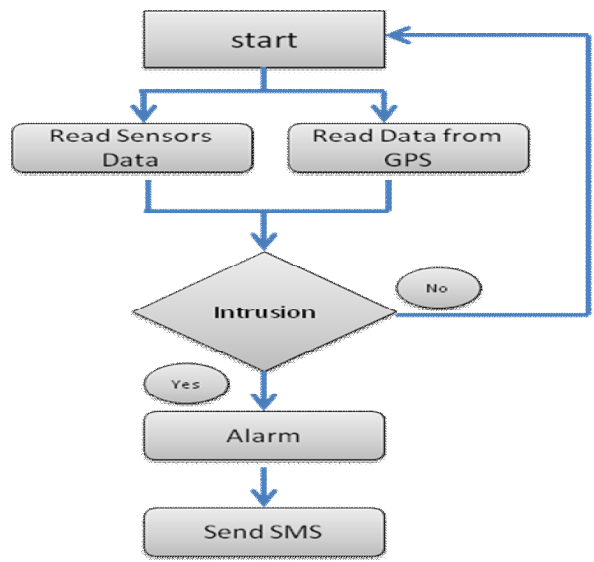

Fig 3: Flow control from unit to client

TABLE 1 : COMMANDS IMPLEMENTED USING THIS SYSTEM

| Command | Description |
| --- | --- |
| AT | Check if the serial interface with GSM modem is working. |
| ATE0 | Turn echo off, less traffic on serial line. |
| AT+CNMI | Display of new incoming SMS. |
| AT+CPMS | Selection of SMS memory. |
| AT+CMGF | SMS string format, how they are compressed. |
| AT+CMGR | Read new message from a given memory location. |
| AT+CMGS | Send message to a given recipient. |
| AT+CMGD | Delete message. |

### IV. PROGRAMMING INTERFACE

Assembly language is being used for the purpose of programming the microcontroller. On intrusion, the numbers to which the message is to be sent and what message is to be sent and in return when the owner of the vehicle sends an SMS then what kind of an action is to be taken on what message, all of it is fed into the microcontroller. The microcontroller takes its decision on the basis of the signals it receives from the mercury switches and the message it receives from the owner of the car [8] [9]. The general flow of the system from system to client and form client to the system has been described through flow charts in figures 3 and 4 respectively.

#### A. Assembly Language

Being a low-level programming language; Assembly language is the best option to program microcontroller. Its each code has one-to-one relation with that of microcontroller's machine level instructions. Rather remembering exact location of data on physical memory, Assembly Language utilizes 'mnemonic codes' or 'symbols'.

#### B. Software interaction with GSM Module SIM300

When microcontroller receives alarm signal from sensor, it require GPS location from Holux-GR89 and send it to SIM300CZ to generate intrusion/theft alert SMS including

### V. BUILDING A GENERIC MODEL AND ITS IMPLEMENTATION

Complete circuit with all hardware was integrated. For our prototype model instead of a big vehicle we used a small car which had doors, bonnet and trunk like other vehicles. The mercury switches/sensors were placed at the bonnet, trunk and doors. After activating the system, when one would try to open bonnet, trunk or door. An alarm would ring and a message will be immediately generated to the owner of the vehicle [10]. In return the owner can take an action, which in our case we locking the gear unit of the vehicle. The system will continuously to trigger a message after every 30 seconds until an action is taken.

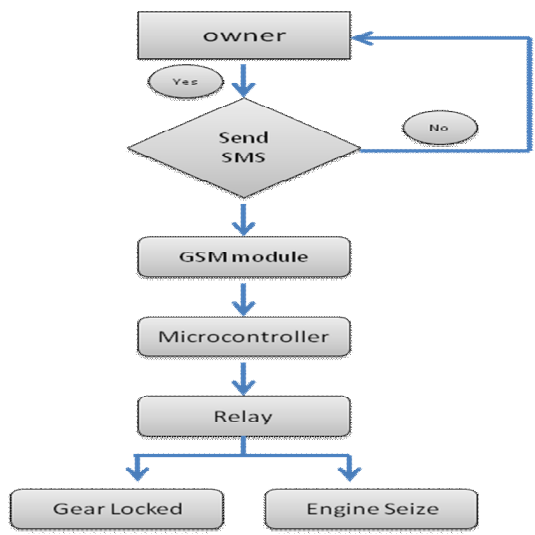

Fig 4: Flow control from client to unit

## VI.  CONCLUSION AND FUTURE WORK

In this project an advance and cost effective approach for vehicle security has been proposed. It can be installed in the vehicle at some hidden place so that it cannot be approached by thieves. GSM and GPS technology has been utilized. It sends alarm signal to the owner of vehicle via SMS, and through SMS vehicle can be jammed. After that system still sends location update SMS after every 30 sec until system is disabled. Proposed system is distinctive in many ways from existing vehicle intrusion and theft control systems; already used systems are either very expensive or ineffective from distance. It is reliable, inexpensive and appropriate design. In future this system can be implemented in all cars, and police numbers can also be added for intrusion notification. For further enhancements sensors with better performance level can be deployed to increase the efficiency and performance level of the system.

ACKNOWLEDGMENT

The authors would like to acknowledge Faculty of Engineering Sciences and Technology (FEST), Iqra University for their persistent support and encouragement throughout the course of the project.